\documentclass[aps,prl,twocolumn,superscriptaddress,amsmath,amsfonts,showpacs]{revtex4}
\usepackage{graphicx}

\begin{document}

%\preprint{}

\title{Topological Order and Non-Abelian Statistics 
in Noncentrosymmetric $s$-Wave Superconductors
}


\author{Satoshi Fujimoto}
\affiliation{Department of Physics, Kyoto University, Kyoto 606-8502, Japan}


%\email[fuji@scphys.kyoto-u.ac.jp] 
%\thanks{}


\date{\today}

\begin{abstract}
We demonstrate that in two-dimensional 
noncentrosymmetric $s$-wave superconductors under applied magnetic fields
for a particular electron density, topological order emerges,
and there exists a zero energy Majorana fermion mode in a vortex core, 
which obeys non-Abelian statistics,
in analogy with $p_x+ip_y$ superconductors,
the Moore-Read Pfaffian quantum Hall state, and
the gapped non-Abelian spin liquid phase of the Kitaev model.
 \end{abstract}



\maketitle

%\section{}
%\subsection{}
%\subsubsection{}

Recently, there is considerable interest in 
emergent topological states of many-body quantum systems 
characterized by a topologically
nontrivial structure of the Hilbert space, i.e. ``topological order''\cite{wen}.
In certain classes of topological states in 2+1 dimension, 
quasiparticles are non-Abelian anyons 
\cite{moore,nayak,fradkin,read,ivanov,stern2,stone1,kitaev,yao}. 
The essential feature of the non-Abelian statistics is that
the exchange of particles is described by unitary operations 
in the multi-dimensional space, which is spanned by the basis of
the degenerate many-body ground state. 
Thus, the state depends on the order of 
the multiple exchange processes of particles.
The possible realization of non-Abelian statistics in real systems
has been extensively studied so far
in connection with 
the $\nu=5/2$ fractional quantum Hall (FQH) state,
and the vortex state of chiral $p_x+ip_y$ 
superconductors (superfluids)
\cite{moore,nayak,fradkin,read,ivanov,stern2,bonderson,stern}.
These topological states are of interest also in the context of
quantum computing, since
the non-Abelian anyon can be utilized as a decoherence-free qubit,
and potentially applied to
the construction of fault-tolerant topological 
quantum computers \cite{freedman,das,stone1,tewari}.
In this paper, 
we present another candidate of a topological phase allowing 
the existence of non-Abelian anyons,
which can be realized in strongly noncentrosymmetric (NC) 
$s$-wave superconductors.
%(superfluids). 
This topological phase belongs to the same class as those
of the Moore-Read (MR) Pfaffian FQH state \cite{moore},
$p_x+ip_y$ superconductors \cite{ivanov},
and the gapped non-Abelian
spin liquid phase of the Kitaev model \cite{kitaev,yao,lee}.
%for which quasiparticles are called Ising anyons.
In NC superconductors, the asymmetric spin-orbit (SO) interaction
which breaks inversion symmetry plays important roles in
various exotic superconducting properties 
\cite{Edelstein,Gorkov,Frigeri,Fujimoto,fuji2}.
In our proposal, the asymmetric SO interaction combined 
with an external magnetic field yields the nontrivial topological state
for a particular electron filling.

We consider type II NC $s$-wave superconductors with 
the Rashba type SO interaction in two dimension. 
We neglect the parity-mixing of triplet components of Cooper pairs
due to the asymmetric SO interaction \cite{Edelstein}, 
because the inclusion of this effect
does not change the essential part of our argument.
For a concreteness, we define our model on the square lattice, though
the following consideration do not reply on 
the particular choice of the crystal structure.
Then the model Hamiltonian is
\begin{eqnarray}
&&\mathcal{H}=\sum_{k,\sigma}\varepsilon_kc^{\dagger}_{k\sigma}c_{k\sigma}
+\alpha\sum_{k,\sigma\sigma'}\mbox{\boldmath $\mathcal{L}$}_0(k)\cdot
\mbox{\boldmath $\sigma$}_{\sigma\sigma'}c^{\dagger}_{k\sigma}c_{k\sigma'} \nonumber \\
&&-\sum_{k}[\Delta c^{\dagger}_{k\uparrow}c^{\dagger}_{-k\downarrow}+h.c.] 
\label{ham1}
\\
&&=\sum_{\nu=\pm}[\sum_{k}\varepsilon_{k\nu}a^{\dagger}_{k\nu}a_{k\nu}
-\sum_{k}\{\Delta_{\nu}(k)a^{\dagger}_{k\nu}a^{\dagger}_{-k\nu}+h.c.\}]
\label{ham2}
\end{eqnarray}
Here, $c^{\dagger}_{k\sigma}$ ($c_{k\sigma}$) is a creation (an annihilation)
operator for an electron with momentum k, spin $\sigma$.
The energy band dispersion is
$\varepsilon_k=-2t(\cos k_x+\cos k_y)-\mu$. 
The second term of eq.(\ref{ham1}) is the Rashba SO interaction with
$\mbox{\boldmath $\mathcal{L}$}_0(k)=(\sin k_y, -\sin k_x, 0)$.
Eq.(\ref{ham2}) is expressed in terms of the chirality basis which diagonalise
the SO term. %; i.e. 
%\begin{eqnarray}
%a^{\dagger}_{k}
%\end{eqnarray}
The energy band is split into two parts by the SO interaction:
$\varepsilon_{k\pm}=\varepsilon_k
\pm\alpha |\mbox{\boldmath $\mathcal{L}$}_0(k)|$.
The gap function in this basis is odd parity, and possesses 
the momentum dependence
$\Delta_{\pm}(k)=\Delta \eta_{\mp}(k)$ with
$\eta_{\pm}(k)=-(\mathcal{L}_{0x}\pm i\mathcal{L}_{0y})
/\sqrt{\mathcal{L}_{0x}^2+\mathcal{L}_{0y}^2}$, which is, importantly,
similar to that of $p_x+ip_y$ superconductors.
%For small $k$, it depends on the polar angle of the momentum $\theta_k$ through
%the phase factor $e^{\pm i\theta_k}$.
Thus, for $\Delta \ll E_F$, we can exploit the same argument 
as that applied to $p$-wave
superconductors \cite{volovik,kopnin}, 
and find that in the mixed state with vortices parallel to 
the $z$-axis, there is
a zero-energy quasiparticle state of a vortex core which is described by
a Majorana fermion.
However, in this case, the existence of Majorana fermions in vortices does not
directly lead to the non-Abelian statistics of them, 
because there are two bands
($\varepsilon_{k\mu}$, $\mu=\pm$), each of which contributes 
to a zero-energy mode with a different band index. 
The existence of two different species of Majorana fermions in a single vortex
implies that the sign change of the fermion operators under the braiding
of two vortices, which is a hallmark of the non-Abelian 
statistics \cite{moore,ivanov}
(see below), is canceled.
%To realize the non-Abelian statistics of vortices, we need to eliminate
To eliminate this unwanted multiplicity of Majorana fermions,
%For this purpose,
we tune the chemical potential as $\mu=-4t$ for which 
the Fermi level crosses the $\Gamma$ point in the Brillouin Zone (BZ). 
In this situation, there are still two bands
near the Fermi level in the model (\ref{ham2}): 
one from $\varepsilon_{k-}$ with a finite Fermi 
momentum, and the other
in the vicinity of the $\Gamma$ point which is given by the Dirac cone.
To generate the mass gap in the Dirac cone,
we introduce the Zeeman coupling 
$\mu_{\rm B}H_z\sum_k(c^{\dagger}_{k\uparrow}c_{k\uparrow}
-c^{\dagger}_{k\downarrow}c_{k\downarrow})$.
%$\mbox{\boldmath $H$}=(0,0,H_z)$.
The magnitude of the gap is of the order $\mu_{\rm B}H_z$.
Then, there is only a single energy band
$\varepsilon_{k-}$ which crosses the Fermi level.
%It is noted that the Zeeman field does not destroy the superconducting state
%even when $\mu_{\rm B}H_z> \Delta$, as long as
%the SO splitting of the energy band is 
%much larger than the Zeeman energy.
Let us assume that $H_z$ is sufficiently smaller than
the orbital depairing field $H_{\rm orb}$.
The Pauli depairing effect due to $H_z$ is negligible
for $\alpha\gg\mu_{\rm B}H_z,\Delta$ \cite{Frigeri}.
%The feasibility of this condition will be discussed in the latter part of 
%this paper.
Under these circumstances, we can integrate out contributions 
from quasiparticles with a gap $\sim\mu_{\rm B}H_z$ at the $\Gamma$ point,
and obtain the low-energy effective Hamiltonian $\mathcal{H}_{\rm eff}$ 
for the single band superconductor.
Taking account of the fact that the Zeeman field $H_z$ induces
the inter-band Cooper pairing between 
the $(+)$-band and the $(-)$-band \cite{fuji2}, we obtain
\begin{eqnarray}
\mathcal{H}_{\rm eff}=\sum_{k}\tilde{\varepsilon}_{k-}a^{\dagger}_{k-}a_{k-}
-\sum_{k}[\tilde{\Delta}_{-}(k)a^{\dagger}_{k-}a^{\dagger}_{-k-}+h.c.].
\label{ham3}
\end{eqnarray}
Here the renormalized energy band is 
$\tilde{\varepsilon}_{k-}=\varepsilon_{k-}+\varepsilon_0$ with
%$\varepsilon_0$ a constant energy shift given by
%\begin{eqnarray}
$\varepsilon_0=
H_z^2\Delta^2/\alpha^2|\mbox{\boldmath $\mathcal{L}$}(k_{F})|^2m_0$,
%\varepsilon_0=
%\frac{H_z^2\Delta^2}{\alpha^2|\mbox{\boldmath $\mathcal{L}$}(k_{F})|^2m_0},
%\end{eqnarray}
%\begin{eqnarray}
$ |\mbox{\boldmath $\mathcal{L}$}(k)|=\sqrt{\mathcal{L}_{0x}^2+
\mathcal{L}_{0y}^2+\mu_{\rm B}^2H_z^2/\alpha^2}$,
%\end{eqnarray}
and $m_0=(4\alpha^2|\mbox{\boldmath $\mathcal{L}$}(k_{F})|^4
+\Delta^2(\mathcal{L}_{0x}^2+\mathcal{L}_{0y}^2))/
2\alpha |\mbox{\boldmath $\mathcal{L}$}(k_{F})|^3$.
$m_0$ is an energy gap of quasiparticles 
in the $(+)$-band in the vicinity
of the Fermi momentum of the $(-)$-band.
The superconducting gap function is 
$\tilde{\Delta}_{-}(k)=a
\Delta\tilde{\eta}_{+}(k)$ where
$a=1+H_z^2\Delta^2/
2\alpha^3|\mbox{\boldmath $\mathcal{L}$}(k_{F})|^3m_0$,
%\begin{eqnarray}
%a=1+\frac{H_z^2\Delta^2}
%{2\alpha^3|\mbox{\boldmath $\mathcal{L}$}(k_{F})|^3m_0},
%\end{eqnarray}
and $\tilde{\eta}_{+}(k)=-(\mathcal{L}_{0x}+i\mathcal{L}_{0y})
/|\mbox{\boldmath $\mathcal{L}$}(k)|$.
The above expression (\ref{ham3}) is valid only in
the vicinity of the Fermi momentum $k_{F}$ defined by 
$\varepsilon_{k_{F}-}=0$ (not $\tilde{\varepsilon}_{k_{F}-}=0$).
In many type II superconductors, it is typical that
for $H_z<H_{\rm orb}$, $\mu_{\rm B}H_z<\Delta$. This implies that
the gap generated by the magnetic field at the $\Gamma$ point might be
smaller than the superconducting gap, which may 
invalidate the approximation
used in the derivation of (\ref{ham3}).
However, our argument on the low-energy vortex core states which
we are most concerned with
is applied only to energy scale $<\Delta^2/E_F$.
Therefore, 
the effective Hamiltonian (\ref{ham3}) is applicable for our purpose,
as long as the condition $\Delta^2/E_F<\mu_{\rm B}H_z<H_{\rm orb}$ is satisfied,
which can be fulfilled in ordinary experimental situations.

The topological order of the model (\ref{ham3}) clearly
manifests in the Chern number which is, for the Hamiltonian of the form
$\mathcal{H}_{\rm eff}=
\sum_{\mu=x,y,z}\sum_k(a_{k-}^{\dagger},a_{-k-})\sigma_{\mu}E_\mu(k)
(a_{k-},a^{\dagger}_{-k-})^t$, defined as \cite{read,lee}
\begin{eqnarray}
\mathcal{N}
=\int \frac{d^2k}{8\pi}\epsilon_{ij}\hat{\mbox{\boldmath $E$}}\cdot
(\frac{\partial \hat{\mbox{\boldmath $E$}}}{\partial k_i}\times 
\frac{\partial \hat{\mbox{\boldmath $E$}}}{\partial k_j}
)
\label{chern}
\end{eqnarray}
where $\hat{\mbox{\boldmath $E$}}=(E_x(k),E_y(k),E_z(k))
/|\mbox{\boldmath $E$}(k)|$.
The integral of (\ref{chern}) is taken over the whole BZ,
while the expression of (\ref{ham3}) is derived for $k$ in the vicinity of
$k_{F}$.
Nevertheless, we can consider the Chern number of the model (\ref{ham3})
by re-interpreting eq.(\ref{ham3}) as a lattice regularized version of
the low-energy effective theory, and extending the $k$-space in which
the model (\ref{ham3}) is defined to the entire BZ.
Then, the numerical evaluation of $\mathcal{N}$ 
for the Hamiltonian (\ref{ham3}) gives
$\mathcal{N}=1$. 
Therefore, the model (\ref{ham3}) is classified as the same topological
class as those of the MR state,
spinless $p_x+ip_y$ superconductors, 
and the gapped non-Abelain phase of the Kitaev model.
The existence of the Zeeman field $H_z$ in the model (\ref{ham3}) 
is important for this topological characterization, because
it does not only break time-reversal symmetry, but also 
ensures the differentiability of $E_{x,y}(k)$ for Eq.(\ref{ham3}) 
which is singular at
$k=0$ for $H_z=0$. 
%flaws the definition of the Chern number of our system.

The Chern number $\mathcal{N}=1$ implies the existence of 
%topological order, and also 
zero-energy Majorana fermion modes
in vortices which obey the non-Abeian statistics, as in the case
of $p_x+ip_y$ superconductors \cite{read,ivanov,lee}.
To demonstrate this, we proceed to 
solve the Bogoliubov de-Gennes (BdG) equations for 
the model (\ref{ham3}) with a single vortex inserted parallel to the $z$-axis.
For simplicity, we switch to the continuum model replacing
the energy band $\varepsilon_k$ of eq.(\ref{ham1}) with
$\varepsilon_k'=k^2/2m-\mu$ and $\mbox{\boldmath $\mathcal{L}$}_0$
with $\mbox{\boldmath $\mathcal{L}$}_0'=(k_y,-k_x,0)$.
Furthermore, we assume that the gap amplitude 
$\Delta(\mbox{\boldmath $r$})$ vanishes 
inside of the vortex core,
and is equal to a constant $\Delta$ outside of the core, and
$\Delta \ll E_F$.
Then, in the vicinity of the Fermi surface, the BdG equations corresponding 
to the model (\ref{ham3}) with a single vortex are
\begin{eqnarray}
\left(\begin{array}{cc}
-i\mbox{\boldmath $v$}_F\cdot\nabla+\varepsilon_0 & 
\Delta_0e^{i\frac{\phi}{2}}
\hat{P}e^{i\frac{\phi}{2}} \\
\Delta_0e^{-i\frac{\phi}{2}}
\hat{P}^{\dagger}e^{-i\frac{\phi}{2}} 
& i\mbox{\boldmath $v$}_F\cdot\nabla-\varepsilon_0
\end{array}
\right)
\Psi=\varepsilon\Psi
\label{bdg}
\end{eqnarray}
where $\Psi^t=(u(\mbox{\boldmath $r$}),
v(\mbox{\boldmath $r$}))$, $\hat{P}=-(\partial_x+i\partial_y)$,
$\hat{P}^{\dagger}=-\hat{P}^{*}$, 
and $\Delta_0=a\Delta(\mbox{\boldmath $r$})
/|\mbox{\boldmath $\mathcal{L}$}(k_F)|$.
The BdG equations (\ref{bdg}) are equivalent to those of spinless 
$p_x+ip_y$ superconductors
except that there are the $\varepsilon_0$-terms in the diagonal components, 
which can be formally absorbed into the shift of the Fermi
momentum $k_F\rightarrow k_F-\varepsilon_0/v_F$.
Thus, the solution of (\ref{bdg}) is given by $\Psi=e^{-i\varepsilon_0/v_F}
\Psi_{p+ip}$ with $\Psi_{p+ip}$ the eigen function of the BdG equations
for spinless $p_x+ip_y$ superconductors, and there exists a zero energy mode
inside the vortex core which is separated from the first excited state by
a gap of energy size $\Delta^2/E_F$ \cite{volovik,kopnin}.
The Bogoliubov quasiparticles for this zero energy state are described by 
a Majorana fermion field
%\begin{eqnarray}
$\gamma=\int d\mbox{\boldmath $r$}[u(\mbox{\boldmath $r$})
a^{\dagger}_{-}(\mbox{\boldmath $r$})
+v(\mbox{\boldmath $r$})
a_{-}(\mbox{\boldmath $r$})],
$
%\end{eqnarray}
since $(v^{*}(\mbox{\boldmath $r$}),u^{*}(\mbox{\boldmath $r$}))=
(u(\mbox{\boldmath $r$}),v(\mbox{\boldmath $r$}))$
for $\varepsilon=0$.
Here $a^{\dagger}_{-}(\mbox{\boldmath $r$})=\sum_ka^{\dagger}_{k-}e^{-ikr}$. 

\begin{figure}
\includegraphics[width=0.95 \columnwidth]{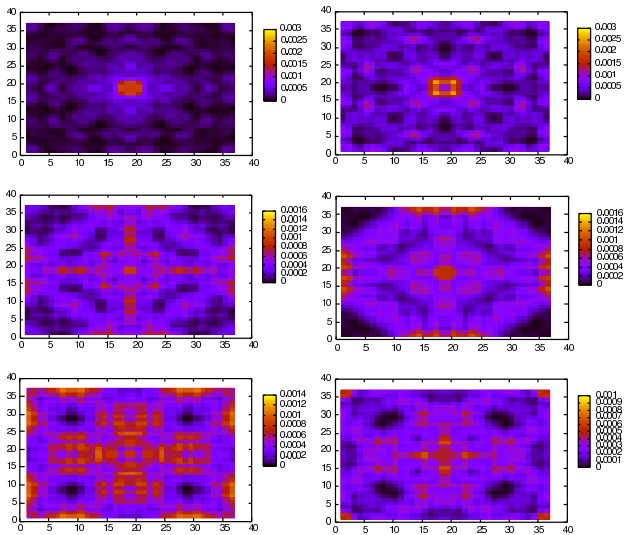}
\caption{\label{Fig01} The density of quasiparticles on the $37\times37$ 
$xy$-plane
for $\varepsilon=4.136\times 10^{-4}t$ (top), 
$\varepsilon=1.289\times 10^{-3}t$ (middle),
and $\varepsilon=1.579\times 10^{-3}t$ (bottom).
The left and right panels are, respectively, 
the plots of $|u(\mbox{\boldmath $r$})|^2$ and 
$|v(\mbox{\boldmath $r$})|^2$. }
\end{figure}

To confirm the above prediction, we apply numerical analysis directly to
the BdG equations for the tight-binding model (\ref{ham1}) 
without referring to the low-energy effective theory (\ref{ham3}).
The energy spectrum and the eigen functions of the BdG equations
were calculated for the model (\ref{ham1}) 
with a vortex located at the center of the system on 
the square lattice with open boundaries.
In this calculation, we assume that the GL parameter is so large that
the Zeeman field $H_z$ is approximated to be uniform, and 
the spatial dependence of the superconducting gap function
due to the vortex is taken into account only in its phase for simplicity.
The topological properties which we are concerned with are not sensitive
to these approximations.
We set parameters as $\mu=-4t$, $\alpha=t$, $\Delta=0.05t$, 
and $\mu_{\rm B}H_z=0.04$.
In FIG.\ref{Fig01} shown are the spatial distributions of the density 
of Bogoliubov quasiparticles for several low-energy states calculated 
for the lattice size $37\times 37$.
The lowest energy state with $\varepsilon=4.136\times 10^{-4}t$ is dominated
by a vortex core state, which can not be
the Caroli-de-Gennes-Matricon mode of the conventional $s$-wave superconductors,
because for our choice of the parameters, the Fermi energy is $E_F=0.25t$, and
$\Delta^2/E_F=0.01t$.  
Also the lowest energy level decreases toward zero 
as the system size increases.
Thus, we identify the lowest energy state with the zero energy mode.
Furthermore, we find the low-energy edge states at the boundaries; e.g. 
for $\varepsilon=1.289\times 10^{-3}$, $1.579\times 10^{-3}$.
The edge state with energy $\ll\Delta$ is a concomitant of 
the zero energy vortex core state,  
which is in accordance with the Chern number 
$\mathcal{N}=1$ \cite{kitaev}. 
%, characterizing the topologically nontrivial state. 
Taking these observations into account,
we can conclude that the NS $s$-wave superconductor with a magnetic field
for the particular electron filling is in the topological state.

As was proved by Ivanov, vortices with zero energy Majorana modes 
obey non-Abelian statistics \cite{ivanov}.
Here we summarize some fundamental properties of 
the non-Abelian statistics relevant to the following argument 
\cite{nayak,ivanov,kitaev,stone1}.
Two Majorana fermion fields $\gamma_i$ and $\gamma_j$ in two vortices 
can be fused into
a complex fermion field $\psi=(\gamma_i+i\gamma_j)/2$.
The fermionic state described by $\psi$ is occupied or unoccupied.
The fusion processes of vortices are analogous 
to the operator product expansions of
the Ising conformal field theory with the central charge $c=1/2$.
%, because of the bulk-edge correspondence for anyonic particles.
The fusion rules for three particle states, a vortex (denoted as $\sigma$), 
a fermion occupied state ($\psi$), 
and an unoccupied vacuum state ($\mbox{\boldmath $1$}$), are
$\sigma\times\sigma=\mbox{\boldmath $1$}+\psi$,
$\psi\times\sigma=\sigma$, and $\psi\times\psi=\mbox{\boldmath $1$}$.
%\begin{eqnarray}
%\sigma\times\sigma=\mbox{\boldmath $1$}+\psi, \quad
%\psi\times\sigma=\sigma, \quad \psi\times\psi=\mbox{\boldmath $1$}.
%\label{fr}
%\end{eqnarray}
The degeneracy of the ground state with $2n$ vortices 
obtained from the fusion rules is $2^{n-1}$.
%,which implies that the quantum dimension of a vortex excitation is 
%$d_{\sigma}=\sqrt{2}$.
%Since the states with different parity of fermion occupation number do not
%mix with each other for energy scale $\ll \Delta$,
%the degeneracy is $2^{n-1}$.
The braiding of vortices is given by a unitary operation 
in this multi-dimensional degenerate space.
%, which is constructed from 
%the spinor representation of $SO(2n)$.
The braiding operator for vortices at $i$ and $j$ is
%\begin{eqnarray}
$R^{\sigma\sigma}=\theta_{\sigma} \exp(\frac{\pi}{4}\gamma_j\gamma_i)
=\theta_{\sigma}\exp(i\frac{\pi}{4}(2\psi^{\dagger}\psi-1))$.
%\label{braid}
%\end{eqnarray}
Here $\theta_{\sigma}$ is a phase factor called a topological spin. 
%which is associated with the conformal spin of the field for a vortex
%$\sigma$. 
Under the exchange of two vortices, the Majorana fermion operators are
transformed as,
$\gamma_i\rightarrow R^{\sigma\sigma}\gamma_{i}(R^{\sigma\sigma})^{\dagger}=\gamma_j$,
$\gamma_i\rightarrow R^{\sigma\sigma}\gamma_{j}(R^{\sigma\sigma})^{\dagger}=-\gamma_i$.
The minus sign in the second transformation rule is essential for
non-Abelian statistics.
The topological spin $\theta_{\sigma}$ in $R^{\sigma\sigma}$ is associated with
the conformal spin of the primary field for vortices which is the chiral 
Ising spin field, 
%with the conformal dimension $(h_0,\bar{h}_0)=(1/16,0)$,
because of the bulk-edge correspondence of anyonic particles:
$\theta_{\sigma}=e^{i2\pi(h_0-\bar{h}_0)}=e^{i\frac{\pi}{8}}$.
As a result, when two vortices are fused into $\psi$, the braiding of them
yields the phase factor $R^{\sigma\sigma}_{\psi}=e^{i\frac{3}{8}\pi}$,
while when the fusion gives the topological charge $\mbox{\boldmath $1$}$,
the phase factor due to the braiding is
$R^{\sigma\sigma}_{1}=e^{-i\frac{\pi}{8}}$.
The braiding rules for the other particle types are $R^{\sigma\psi}=-i$ and
$R^{\psi\psi}=-1$.
%It is useful to write down here 
%other braiding rules which will be used later.
%The phase factor caused by the braiding of $\psi$- and $\sigma$-particles 
%is $R^{\sigma\psi}=-i$, and that by the braiding of two $\psi$-particles is
%$R^{\psi\psi}=-1$.
%Since the fusion outcome of two $\sigma$'s is
%$\mbox{\boldmath $1$}$ or $\psi$,
%a single $\sigma$-particle has the quantum dimension 
%$d_{\sigma}=\sqrt{2}$.
%This nonlocal character implies
%the possible utility
%of the $\sigma$-particles
%as qubits in the context of quantum computing.
%The topological stability of the $\sigma$-particles against any local perturbations
%ensures that the quantum computer constructed from them
%is decoherence-free and fault-tolerant.

\begin{figure}
\includegraphics[width=0.95 \columnwidth]{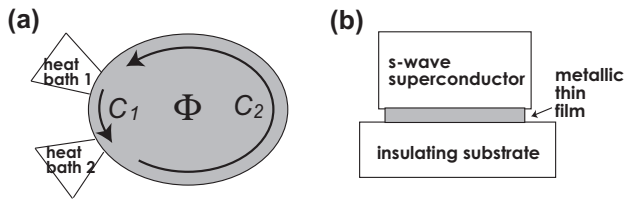}
\caption{\label{Fig02} (a) Experimental setup for the
thermal transport measuremnt of the edge state. 
A magnetic flux $\Phi$
penetrates in the bulk. Two heat baths are attached at
the boundary so that the longer path between the two heat baths, $C_2$, encircles
almost all vortices in the bulk. 
(b) Superconductor-metal-insulator junction. 
%A thin film has
%a thckness smaller than the mean free path of electrons.
}
\end{figure}

We, now, discuss the feasibility of the experimental detection
of the non-Abelian statistics.
One promising approach is to use the two-point-contact interferometer
proposed in the context of the FQH state \cite{fradkin,bonderson,stern}.
In the superconducting state considered here, 
this experiment is applicable only to
the thermal transport.
According to refs. \cite{bonderson,stern}, the interference term of the
edge heat current $J^{\rm int}$
depends on the parity of the total number of vortices $n$ in the bulk.
$J^{\rm int}$ for odd $n$ is much smaller than $J^{\rm int}$ for even $n$,
though both of them do not exhibit 
the dependence on a magnetic flux $\Phi$, because the $\sigma$-particle
is neutral. This parity dependence characterizes the non-Abelian statistics.
%One promising approach is the transport measurement which probes
%interference effects caused by the braiding of the $\sigma$-particles as
%proposed in the context of the FQH state \cite{fradkin,bonderson,stern}.
%In our case, the system is in the superconducting state, and thus
%the transport properties of quasiparticles which should be distinguished from
%contributions of supercurrents appear in the thermal conductivity.
%The interference effect due to the braiding of
%non-Abelian anyons can be observed in the heat current carried by
%the edge quasiparticles.
%Since the $\sigma$-particle is neutral, 
%in order to introduce the interaction between edge quasiparticles and
%an Aharonov-Bohm flux $\Phi$,
%We consider a system slightly different from that proposed 
%in refs. \cite{fradkin,bonderson,stern}.
%It consists of a NC $s$-wave superconductor with a finite closed boundary and 
%two constrictions which separate the system into three regions, I, II, and III,
%as depicted in FIG.\ref{Fig02}.
%A magnetic flux $\Phi$ is penetrating in the region II.
%At the constrictions, tunneling of edge quasipaticles between opposite sides
%of the boundary is possible. 
%Two heat baths are attatched to the boundary of the region I.
%The temperature difference between them
%must be sufficiently smaller than $\Delta^2/E_F$.
Another possible experiment is a bit indirect but simpler.
It uses a disk-shaped 
system with which two heat baths are attached at the boundary. 
(see FIG.\ref{Fig02}(a).)
%The chiral edge current flow is anticlockwise.
For this geometry, 
as in the case of $p_x+ip_y$ superconductors \cite{stone2},
the energy spectrum of the edge state %is discrete, and
%, and labeled by the angular momentum.
%Because of the similarity between the model (\ref{ham3}) 
%and $p_x+ip_y$ superconductors,
%we can apply the same analysis of the edge mode 
%as that for $p_x+ip_y$ superconductors to
%our problem, which indicates that 
depends on the parity of the total number of vortices $n$
in the bulk. % \cite{stone2}.
For even $n$, the lowest energy state has 
a gap of the order $\Delta/k_FL$ where $L$ is the length of the boundary.
Although the gap is small for a sufficiently large system size,
it is nonzero, and thus the quasiparticle corresponding to this edge mode is
a complex fermion interacting with $\Phi$. 
This Bogoliubov quasiparticle is categorized
as the same particle type as the $\psi$-fermion in the bulk, because,
in the limit that the two vortices merge together at a position 
$\mbox{\boldmath $r$}$,
the resulting $\psi$-particle is nothing but  
the Bogoliubov quasiparticle with a nonzero energy \cite{stone1}.
For odd $n$, the low-energy edge state is
a Majorana fermion mode, and can be fused with an unpaired Majorana fermion
in the bulk resulting in the $\psi$-state or 
the $\mbox{\boldmath $1$}$-state.
The phase accumulated by the current flow of the edge $\sigma$- or 
$\psi$-particles
encircling the bulk $n$ vortices is obtained 
from the square of the braiding operator $(R^{ab})^2$ ($a,b=\sigma,\psi$).
%The phase accumulated by the current flow of the edge $\sigma$- or 
%$\psi$-particles
%encircling the bulk $n$ vortices in the region II is obtained 
%from the square of the braiding operator 
%$(R^{ab})^2$ ($a,b=\sigma,\psi$).
When the temperature $T_1$ of the heat bath $1$ is smaller than
the temperature $T_2$ of the heat bath $2$, the chiral edge heat current 
flows mainly in the path $C_2$ (anticlockwise direction) encircling
the bulk vortices. 
In this case, 
for even $n$, the edge heat current carried by the $\psi$-particles
exhibits a usual dependence on $\Phi$, i.e. 
$J^{\rm int}_{\rm even}\sim \sum_{m=1}^{\infty}A_m\cos(2\pi m e\Phi/hc)$,
while, for odd $n$, using the fusion rules and 
the braiding rules mentioned above, we obtain
$J^{\rm int}_{\rm odd}\sim \sum_{m=1}^{\infty}B_{4m}\cos \pi m$, where
the $B_{4m}$-term corresponds to a trajectory winding around the boundary loop
$4m$ times. Thus $J^{\rm int}_{\rm odd}$ is much suppressed.
On the other hand,
when $T_1>T_2$, the edge current flows mainly in the path $C_1$, 
less affected by $\Phi$. In this case, the dependence of $J^{\rm int}$
on the parity of $n$ is weaker than the case of $T_1<T_2$.
%By using the fusion rules and the braiding rules mentioned above, 
%the interference term of the heat current reflecting
%these phase changes is calculated as
%$J^{\rm inter}_{\rm odd}\sim\sum_{m=1}^{\infty}
%A_m\cos m\frac{\pi}{2}$ for odd $n$. Here $A_m$ is a constant, corresponding to
%a trajectory winding around the boundary of the II region $2m$ times.
%$J^{\rm inter}_{\rm odd}$ is vanishingly small, as was pointed out 
%in refs. \cite{bonderson,stern}.
%For even $n$, we have
%%\begin{eqnarray}
%$J^{\rm inter}_{\rm even}\sim\sum_{m=1}^{\infty}
%B_m\cos m\phi. $
%\end{eqnarray}
%Here $\phi=2\pi e\Phi/hc$, and $B_m$ is a constant.
%Therefore, the difference of the interference effect between 
%the odd and even $n$ cases
%characterizes the non-Abelian statistics.
%, as in the case of the FQH state.
These observable effects can be utililzed for the detection of the non-Abelian
statistics.

An advantage of NC $s$-wave superconductors over $p_x+ip_y$-wave superconductors
and the $\nu=5/2$ FQH state 
is that the gap energy scale of the former can be typically much larger than
those of the latter. 
Note that the superconductivity in NC systems needs not to be
a bulk phenomenon.  
Let us consider the junction between an $s$-wave superconductor and
a metallic thin film placed on an insulating substrate. 
(see FIG.2(b).) 
The thin film must be sufficiently clean so that 
the mean free path is larger than its thickness.
In this system, inversion symmetry is broken, and
an asymmetric potential gradient perpendicular to
the interface is introduced.
We can use a material with 
a high transition temperature such as 
MgB$_2$ ($T_c\sim 39$K) for the superconductor \cite{aki}.
Then, the proximity effect induces $s$-wave superconductivity 
in the 2D NC system realized in the thin film.
If the Fermi energy of the film $E_F$ is much smaller than
that of the bulk superconductor, the energy gap in the vortex core 
$\Delta^2/E_F$ 
for the proximity-induced NC superconductor can be relatively large.
The strength of the asymmetric SO interaction can be controlled by 
changing the substrate or applying a perpendicular
voltage on the film.
Although electrons should experience strong SO scatterings at the interface,
the transition temperature and the gap of the $s$-wave pairing state
are not affected by them.
Also, note that the Majorana fermions in vortices of 
the NC superconductors
do not require a half quantum vortex, i.e. a texture of the 
$\mbox{\boldmath $d$}$-vector, because our system is essentially regarded as
spinless. 
%(In other words, the spin degrees of freedom is 
%quenched by the SO interaction.) 
In this sense, zero energy Majorana states in
NC $s$-wave superconductors are more realizable than
in spinful $p$-wave superconductors.

In conclusion, 
NC $s$-wave superconductors under magnetic field
have a topological order for a particular electron filling, and can be
playgrounds for the non-Abelian anyons. 
Although we consider only the Rashba SO interaction here,
our argument can be easily generalized 
to other asymmetric SO interactions.

The author thanks N. Kawakami and R. Ikeda 
for invaluable discussions.
The numerical calculations were performed on SX8 at YITP
in Kyoto University.
This work was supported by a Grant-in-Aid from the Ministry
of Education, Science, Sports and Culture, Japan.

%\bibliography{ZZ}

\end{document}